\documentclass[conference, a4paper]{IEEEtran}

\usepackage[utf8]{inputenc} 
\usepackage[T1]{fontenc}
\usepackage{url}
\usepackage{ifthen}
\usepackage{cite}

 \usepackage[dvipdfmx]{graphicx}
 \usepackage{grffile}
 \usepackage{algorithm}
 \usepackage{algpseudocode}
\makeatletter
\newcommand{\removelatexerror}{\let\@latex@error\@gobble}
\makeatother
\algnewcommand\algorithmicinput{\textbf{Input:}}
\algnewcommand\INPUT{\item[\algorithmicinput]}
\algnewcommand\algorithmicoutput{\textbf{Output:}}
\algnewcommand\OUTPUT{\item[\algorithmicoutput]}

\usepackage{color}

\usepackage{amssymb}
\usepackage[cmex10]{amsmath} 

\allowdisplaybreaks[1]

\newcommand{\F}[0]{{\mathbb F}}

\begin{document}
\title{Asymptotic Analysis on Spatial Coupling Coding for Two-Way Relay Channels} 


\author{%
   \IEEEauthorblockN{	Satoshi Takabe\IEEEauthorrefmark{1},
   			Yuta Ishimatsu\IEEEauthorrefmark{1},
                        Tadashi Wadayama\IEEEauthorrefmark{1}, and 
                        Masahito Hayashi\IEEEauthorrefmark{2}\IEEEauthorrefmark{3}}
  \IEEEauthorblockA{\IEEEauthorrefmark{1}
  	Department of Computer Science, Faculty of Engineering, Nagoya Institute of Technology}
  \IEEEauthorblockA{\IEEEauthorrefmark{2}
  	 Graduate School of Mathematics, Nagoya University}
  \IEEEauthorblockA{\IEEEauthorrefmark{3}
  	 Centre for Quantum Technologies, National University of Singapore}
  \IEEEauthorblockA{
  	Email: s\_takabe@nitech.ac.jp, 26115016@stn.nitech.ac.jp, 
        wadayama@nitech.ac.jp, masahito@math.nagoya-u.ac.jp}
                    
}


\maketitle

\begin{abstract}
Compute-and-forward relaying is effective to increase bandwidth efficiency 
of wireless two-way relay channels.
In a compute-and-forward scheme, a relay tries to decode a linear combination 
composed of transmitted messages from other terminals or relays.
Design for error correcting codes and its decoding algorithms 
suitable for compute-and-forward relaying schemes are still important issue to be studied.
In this paper, we will present an asymptotic performance analysis on LDPC codes 
over two-way relay channels based on density evolution (DE).  
Because of the asymmetric nature of the channel, 
we employ the population dynamics DE combined with DE formulas for asymmetric channels to obtain BP thresholds.
In addition, we also evaluate the asymptotic performance of spatially coupled LDPC codes for two-way relay channels.
The results indicate that the spatial coupling codes yield improvements in the BP threshold compared with 
corresponding uncoupled codes for two-way relay channels.
\end{abstract}

\section{Introduction}

A relaying with an appropriate signal processing and decoding are ubiquitous in wireless communications
such as satellite communications, mobile wireless communications,  and wireless local area networks.
Increasing demand for band width efficiency in wireless communications
promotes spread of research activities on relaying and forwarding techniques.
For example, theoretical limits of efficiencies of relaying techniques 
such as decode-and-forward \cite{Cover79} and amplify-and-forward \cite{Laneman} have been deeply studied.
Recently, Nazar and Gastpar presented a novel concept of {\em compute-and-forward relaying} \cite{Nazer11}.
In a compute-and-forward scheme, a relay tries to decode a linear combination 
composed of transmitted messages from other terminals (or relays)
and then the relay forwards a decoded linear combination to 
another relay or a terminal. That is, the repeater has no intention to decode each message
separately. 
The concept is also termed as wireless network coding 
or {\em physical layer network coding}
 that has invoked huge research interests \cite{Katti08, Zhang09}.
Recently, Sula \cite{Sula17} \textit{et al.} presented a practical decoding scheme for LDPC codes 
in compute-forward multiple access (CFMA) systems.
Ullah et al. \cite{Ullah17} derived the random coding error exponent for 
the uplink phase of a two-way relay channel.

The simplest scenario for a compute-and-forward scheme may be 
{\em wireless two-way relay channels} \cite{Narayanan}.
Two terminals A,  B and a relay R are involved in this channel.
The terminal A has own message and wishes to send 
it to the terminal B. Similarly, the terminal B wishes to send 
own message to A. There is no direct 
wireless connection between A and B, but a relay R has 
bi-directional wireless connections to both of A and B.
When the relay R can  decode a linear combination successfully, 
it is broadcasted to A and B in the next time slot.
The terminals A and B can recover an intended message 
by subtracting own message from the received message.

In order to obtain a highly reliable estimate of linear combination at the relay, 
appropriate error correcting codes should be exploited because
the received signal is distorted by additive noises.
In such a case, the relay R intends to {\em decode} a sum of two codewords 
sent from A and B.  One possible candidate of error correcting codes 
for such a situation is {\em low-density parity-check (LDPC) codes} \cite{Gallager63}.
A combination of LDPC codes and belief propagation (BP) decoding 
has been proved to be very powerful and effective for additive noise channels \cite{MacKay99}.
Sula et al.  \cite{Sula17} discussed an appropriate modified BP decoding for 
the two-way relay channels. They presented a performance analysis on 
LDPC codes over two-way relay channel based on computer simulations.

The goal of this work is twofold. The first goal is to provide 
an asymptotic performance analysis for LDPC codes over two-way relay channels
based on {\em density evolution} (DE). DE \cite{Richardson} is a common theoretical tool 
to study the asymptotic typical behavior of a BP decoder and it provides BP thresholds of the
target channel. Although the BP threshold is below the Shannon limit, a BP threshold 
indicates a practical achievable rate with low complexity encoding and decoding.
One technical challenge for evaluating the BP threshold of two-way relay channels comes
from an asymmetric nature of the channel. That is, we cannot rely on the zero codeword 
assumption commonly used in DE analysis for binary-input memoryless output-symmetric channels \cite{Richardson}.
In order to overcome this difficulty, we will employ population dynamics DE \cite{Mezard}
combined with the DE formula derived by Wang et al.  for asymmetric channels \cite{Wang05}.

The second goal of this paper is to provide DE analysis for spatially coupled LDPC (SC-LDPC) codes over two-way relay channels.
It is known that appropriately designed spatially coupled codes yield improvements in BP thresholds compared with 
those of uncoupled regular LDPC  codes with comparable parameters \cite{Zigangirov99} \cite{Lentmaier10}.
In many cases, we can observe {\em threshold saturation} \cite{Kudekar13}, i.e., a phenomenon 
that the BP threshold converges to the MAP threshold.
The same is true for the spatially coupling coding for two-way erasure multiple access channels
for a joint compute-and-forward scheme~\cite{Hern}.
As far as the authors know, typical behavior of BP decoding of 
spatially coupled LDPC codes over 
the two-way relay channels except for erasure ones is unknown. 
We consider that it is worth studying not only from practical interests but also from theoretical interests
to provide an example of the DE analysis for general asymmetric channels.
In this work, we will extend
the population dynamics DE to protograph codes~\cite{Thorpe04} and  
perform numerical evaluations. 
Recent work by Hayashi et al.~\cite{Hayashi} shows that 
efficient codes for two-way relay channels are useful
to establish secure communication with untrusted relay.
The spatially coupled codes presented here can be regarded as promising candidates for such uses.

\section{Preliminaries}

\subsection{Problem setting}

The wireless channel model assumed here is described as follows.
Let $X_A^{(t)}$ (resp. $X_B^{(t)}$) be a binary random variable where
$t = 1,2, \ldots$ represents time index.
The binary-bipolar 
conversion function $\mu: \{0,1\} \rightarrow \{+1, -1\}$,
$\mu(x) = 1 - 2 x$ is applied to $X_A^{(t)}$ and $X_B^{(t)}$ 
before their transmission. This means that we assume 
binary phase shift keying (BPSK) as modulation format.
The terminals A and B then transmits the modulated signals 
$\mu(X_A^{(t)})$ and $\mu(X_B^{(t)})$ to the air. The relay R observes 
a received symbol 
\begin{equation} \label{channel_model}
	Y^{(t)} = \mu(X_A^{(t)}) + \mu(X_B^{(t)}) + W^{(t)}, 
\end{equation}
where $W^{(t)}$ is a zero mean Gaussian random variable with variance $\sigma^2$.
The channel model (\ref{channel_model}) is justified 
under the assumption such that 
perfect symbol/phase synchronization and perfect power control 
are achieved at R.
The relay R tries to infer $X_A^{(t)} \oplus X_B^{(t)}$ 
 from $Y^{(t)}$ 
as correct as possible, where the operator $\oplus$ represents the addition over $\F_2$.  
The channel model is called the {\em two-way relay channel} in this paper.

If no error correcting code is used, then symbol by symbol estimation can be applied.
In the next phase, 
the estimate $\hat x_A^{(t)} \oplus \hat x_B^{(t)}$
is then broadcasted to A and B. 
If the estimate $\hat x_A^{(t)} \oplus \hat x_B^{(t)}$ equals to 
the true value $x_A^{(t)} \oplus x_B^{(t)}$, then 
the terminal A (resp. B) can retrieve $x_B^{(t)}$ (resp. $x_A^{(t)}$) 
from $\hat x_A^{(t)} \oplus \hat x_B^{(t)}$.
This protocol, i.e., wireless network coding \cite{Katti08} \cite{Narayanan},  can be seen as the simplest case of 
the compute-and-forward technique \cite{Nazer11}, and it 
increases bandwidth efficiency of the two-way relay channel.

\subsection{LDPC coding}

Let $C \subset \F_2^n$ be an LDPC code used in 
terminals A and B. The terminals A and B 
independently select own codewords 
$\mathbf{x}_A = (x_{A,1},\ldots, x_{A, N})  \in C$
and $\mathbf{x}_B = (x_{B,1},\ldots, x_{B, N}) \in C$ 
according to their own message. From the channel model 
(\ref{channel_model}), the received word is given by
\begin{equation}
	\mathbf{y} 
	=  (\mu(x_{A,1}),\ldots, \mu(x_{A, N})) + (\mu(x_{B,1}),\ldots, \mu(x_{B, N})) 
	+ \mathbf{w},
\end{equation}
where $\mathbf{w}$ represents additive white Gaussian noise vector.
A decoder, possibly a BP decoder,  tries to recover 
$\mathbf{x}_A \oplus \mathbf{x}_B$ from the received word $\mathbf{y}$.
In this paper, we focus on decoding methods for recovering $\mathbf{x}_A \oplus \mathbf{x}_B$.

\subsection{IID assumption-based belief propagation}

Assume that two stochastic processes  
\[
\{X_A^{(1)}, X_A^{(2)}, \ldots, X_A^{(t)}, \ldots \}, 
\{X_B^{(1)}, X_B^{(2)}, \ldots, X_B^{(t)}, \ldots \}
\]
are IID and that $X_A^{(t)}$ and $X_B^{(t)}$ are independent.
For simplicity, we here assume that $\mathrm{Pr}[X_A^{(t)} = 1] = \mathrm{Pr}[X_B^{(t)} = 1] = 1/2$ holds for 
any $t$. From these assumptions, we have the probability of events:
\begin{eqnarray}
\mathrm{Pr} [\mu(X_A^{(t)}) + \mu(X_B^{(t)}) = 0] &=& 1/2,	\\
\mathrm{Pr} [\mu(X_A^{(t)}) + \mu(X_B^{(t)}) = -2] &=& 1/4,	\\
\mathrm{Pr} [\mu(X_A^{(t)}) + \mu(X_B^{(t)}) = +2] &=& 1/4.	
\end{eqnarray}
Let $Z^{(t)} = X_A^{(t)} \oplus X_B^{(t)}$. From the IID assumption,
$Z^{(t)}$ is also a memoryless stochastic process.
We now consider a {\em virtual channel} whose 
input and output symbols are $Z^{(t)}$ and $Y^{(t)}$, respectively.
It is evident that the prior probability of $Z^{(t)}$ is given by $\mathrm{Pr}(Z^{(t)} = 0) = \mathrm{Pr}(Z^{(t)} = 1) = 1/2$.
Under the IID assumptions,  
the conditional PDF representing the channel statistics 
of the virtual channel is given by 
\begin{equation} 
\begin{aligned}
\!\mathrm{Pr} [Y^{(t)} \!&=\! y | Z^{(t)} \!= 1] \!=\! F(y; 0, \sigma^2), \\ 
\!\mathrm{Pr} [Y^{(t)} \!&=\! y | Z^{(t)} \!= 0] \!=\! \frac{1}{2} F(y; -2, \sigma^2) \!+\!\frac{1}{2} F(y; +2, \sigma^2), 
\end{aligned}\label{vir}
\end{equation}
where $F(y; m, \sigma^2)$ is the Gaussian distribution with mean $m$ and variance $\sigma^2$
defined by
\[
F(y; m, \sigma^2) = \frac{1}{\sqrt{2 \pi \sigma^2}} \exp \left( \frac{-(y-m)^2}{2 \sigma^2} \right).
\]

From this conditional PDF, symbol log likelihood ratio (LLR) can be easily derived:
\begin{equation} \label{llr}
\lambda^{(t)}(y) = \ln \frac{\mathrm{Pr} [Y^{(t)} = y | Z^{(t)} = 0]}{\mathrm{Pr} [Y^{(t)} = y | Z^{(t)} = 1]}
= \ln \left[ \cosh \frac{2y}{\sigma^2} \right] - \frac{2}{\sigma^2}.\
\end{equation}
If the IID assumption is valid, we can make the best estimation on 
$Z^{(t)}$ only from $\lambda^{(t)}$.
Note that this LLR expression is a special case of the 
LLR expression derived by Sula et al. \cite{Sula17}.


Let us go back to the argument on the case where terminals A and B 
employ a binary linear code $C$.
Due to linearity of the code $C$,  it is clear that 
$(Z^{(1)},\ldots, Z^{(n)})$ also belongs to $C$.
From this fact, {\em IID-assumption based maximum likelihood (ML)
decoding} can be defined as
\begin{equation} \label{iidml}
(\hat z_1, \ldots, \hat z_n) = \arg \max_{(z_1,\ldots, z_n) \in C}
\prod_{t = 1}^ n L(y_t | z_t), 
\end{equation}
where the likelihood functions are defined by
\begin{equation}
\begin{aligned}
L[y|1] &= F(y; 0, \sigma^2), \\
L[y|0] &= \frac{1}{2} F(y; -2, \sigma^2) +\frac{1}{2} F(y; +2, \sigma^2).
\end{aligned}
\end{equation}

This ML rule is sub-optimal because the likelihood is based on the IID assumption.
Regardless of its sub-optimality,  the IID assumption makes 
the structure of a decoder simple, and it also makes easier to exploit known channel 
coding techniques developed for memoryless channels.

Belief propagation (BP) decoding for LDPC codes can be regarded as an 
approximation of ML decoding as a message passing form. It would be 
natural to develop a BP decoding algorithm 
for the binary compute-and-forward channel based on the IID assumption-based ML rule (\ref{iidml}).
It is not hard to see that the {\em IID assumption-based BP} coincides with the conventional 
log-domain BP algorithm \cite{Richardson} with symbol LLR expression (\ref{llr}).
This type of BP decoder has already discussed in \cite{Sula17} \cite{Ullah17}.
A significant advantage of the IID assumption-based BP is that it can be easily implemented based on 
a practical BP decoder for the additive white Gaussian noise (AWGN) channel just by 
replacing an LLR computation unit.

%

\section{Density evolution for binary two-way relay channels with IID assumption}

We employ DE to study BP thresholds of binary two-way relay channels with the IID assumption.
In this section, we first introduce the population dynamics DE and estimate the BP threshold for uncoupled regular LDPC codes.
The BP threshold for SC-LDPC is then evaluated.

\subsection{Density evolution for asymmetric channels}

For simplicity, we here focus on  $(d_l,d_r)$-regular LDPC codes, where $d_l$ and $d_r$ represent 
the variable and check node degrees, respectively. Extension to irregular codes is straightforward.
It is noteworthy that we need to handle signal dependent 
noises~(\ref{vir})
for two-way relay channels with the IID assumption. 
This means that we cannot rely on the zero code assumption in a DE analysis. 
In the following, we follow the Wang's DE formulation \cite{Wang05} to overcome this difficulty.

The conditional PDF $P^{(l)}(m|z)$ (resp. $Q^{(l)}(\hat{m}|z)$) denote 
the PDF of a message $m$ from a variable node to a check node
(resp. $\hat{m}$ from a check node to a variable node) with transmitted word $z$ at the $l$-th step.
The distribution of LLR of the virtual channel is denoted by $P^{(0)}(z)$. 
Note that those PDFs depend on a transmitted word because of the asymmetric nature of the channel. 
For symmetric channels, in contrast, the zero code assumption omits the dependence.
Let $\Gamma(P_A)\triangleq P_A\circ \gamma^{-1}$ be a density transformation for a random variable $A$ with distribution $P_A$~\cite{Wang05}
where $\gamma:\mathbb{R}\rightarrow \{0,1\}\times [0,\infty)$, 
$\gamma(m) \!=\!  \left(1_{m\le 0},\ln\coth\left|{m}/{2}\right|\right)$ 
with an indicator function $1_{\{\,\cdot\,\}}$.

The DE equations for binary asymmetric channels~\cite{Wang05} are given by
\begin{align}
P^{(l)}(m|z)&\!=\!P^{(0)}(z)\otimes \left(Q^{(l-1)}(\hat{m}|z\right)^{\!\otimes(d_l-1)}, \label{eq_d1}\\
Q^{(l)}(\hat{m}|z)&\!=\!\Gamma^{-1}\left(\left\{\Gamma\!\left(\frac{P^{(l)}({m}|0)\!+\!P^{(l)}({m}|1)}{2}\right)\right\}^{\!\otimes(d_r-1)}\right.\nonumber\\
&\!\left.+(-1)^z\!\left\{\Gamma\!\left(\frac{P^{(l)}({m}|0)\!-\!P^{(l)}({m}|1)}{2}\right)\right\}^{\!\otimes(d_r-1)}\right), \label{eq_d2}
\end{align}
where $\otimes$ denotes the convolution operator on PDFs.
Although these convolutions of PDFs can be efficiently evaluated with fast Fourier transformation, 
numerical evaluation requires huge computational costs.
We use an alternative approach, population dynamics \cite{Mezard}, to reduce computational complexity because 
the DE analysis for SC-LDPC codes 
 deals with a number of DE equations simultaneously.

Equations (\ref{eq_d1}) and~(\ref{eq_d2}) have 
equivalent forms called the replica-symmetric cavity equations~\cite{Mezard}, which read
\begin{align}
P^{(l)}(m|z)&\!=\!\int dyL[y|z]\int \prod_{s=1}^{d_l-1}d\hat{m}^{(s)}Q^{(l-1)}(\hat{m}^{(s)}|z)\nonumber\\
&\times\delta\left(m-\lambda(y)-\sum_{s=1}^{d_l-1}\hat{m}^{(s)}\right), \label{eq_d3}\\
Q^{(l)}(\hat{m}|z)&\!=\!\frac{1}{2^{d_r-2}}
\sum_{\{z^{(s)}\}\in S}
\int \prod_{s=1}^{d_r-1}d{m}^{(s)}P^{(l)}({m}^{(s)}|z^{(s)})\nonumber\\
&\times\!\delta\left(\hat{m}\!-\!2\tanh^{-1}\left[\prod_{s=1}^{d_r-1}\tanh\left(\frac{{m}^{(s)}}{2}\right)\right]\right),\! \label{eq_d4}
\end{align}
where $\lambda(y)$ denotes the LLR defined as the r.h.s. of (\ref{llr}) and 
$S\triangleq\left\{\{z^{(s)}\}\in \{0,1\}^{d_r};\bigoplus_{s=1}^{d_r}z^{(s)}=0, z^{(d_r)}=z\right\}$.

In Algorithm 1, we describe a procedure of the population dynamics DE.
In population dynamics, the PDFs $P(\cdot|z)$ and $Q(\cdot|z)$ ($z\in\{0,1\}$) are approximated
to histograms (populations) of $N$ samples denoted by, e.g., $\{\nu_i^0\}$ ($i\in [N] \triangleq \{1,\dots,N\}$).
The parameter $N$ is called the population size and 
the DE equations are exactly solved in the large-$N$ limit.
Each sample is recursively updated by an update rule written in a delta function $\delta(\cdot)$ 
in~(\ref{eq_d3}) or~(\ref{eq_d4}).
After each iteration finishes, we can estimate bit error rate (BER) at the step.
Although the recursion should continue until every population converges,
it stops at the maximum iteration step $T$ in practice.

\begin{figure}[!t]\label{alg}
 \removelatexerror
  \begin{algorithm}[H]
   \caption{Population Dynamics DE}
  \begin{algorithmic}[1]
   \INPUT Population size $N$, Maximum iteration $T$ 
   \OUTPUT Populations $\{\nu_i^0\}$, $\{\nu_i^1\}$, $\{\hat{\nu}_i^0\}$, and $\{\hat{\nu}_i^1\}$ ($i\in[N]$)
   \State Initialization: $\nu_i^0=\nu_i^1=0$
   \For{$l= 1$ to $T$} 
      \For{$z=0$ to $1$} \Comment{Update of $\{\hat{\nu}_i^z\}$ ($Q(\cdot|z)$)}
        \For{$i= 1$ to $N$}
          \State{Draw $z(1),\dots,z(d_r-1)$ uniformly in $\{0,1\}$ to satisfy $z\oplus \left(\bigoplus_{s=1}^{d_r-1}z(s)\right)=0$.}
      	  \State{Draw $i(1),\dots,i(d_r-1)$ uniformly in $[N]$.}
          \State{$\hat{\nu}_i^z \leftarrow 2\tanh^{-1}\left[\prod_{s=1}^{d_r-1}\tanh\left({\nu}_{i(s)}^{z(s)}/2\right)\right]$.}
        \EndFor
      \EndFor
      \For{$z= 0$ to $1$} \Comment{Update of $\{{\nu}_i^z\}$ ($P(\cdot|z)$)}
        \For{$i= 0$ to $N$} 
 	  \State{Draw $y$ {from} $L[y|z]$.}
      	  \State{Draw $i(1),\dots,i(d_l-1)$ uniformly in $[N]$.}
          \State{${\nu}_i^z \leftarrow \lambda(y)+\sum_{s=1}^{d_l-1}\hat{\nu}_{i(s)}^{z}$.}
        \EndFor
      \EndFor
   \EndFor
  \end{algorithmic}
  \end{algorithm}
\end{figure}

\begin{figure}[!t]
\centering
\includegraphics[width=0.87\linewidth]{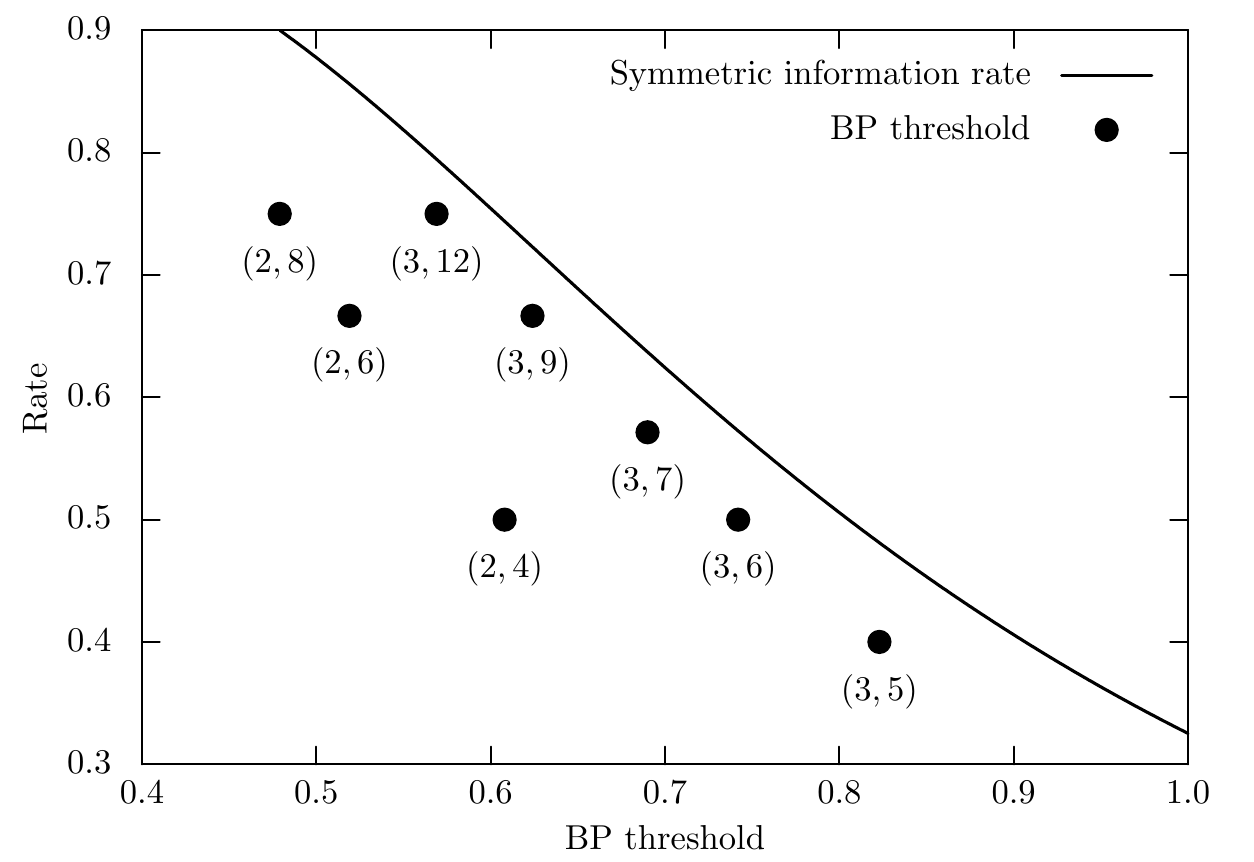}
\caption{BP thresholds for regular LDPC code ensembles over two-way relay channels  versus code rate.
The solid line represents the symmetric information rate of the channel.
}\label{zu_s1}
\end{figure}

We evaluate a BP threshold defined as a threshold of $\sigma$ in the virtual channel~(\ref{vir}) below which
 LDPC codes are typically decodable by a BP decoder.
As a MAP threshold, we use the {\em symmetric information rate} $\sigma_{sym}(R)$
defined as a solution of $C_{sym}(\sigma_{sym}(R))=R$ for code rate $R$, where
\begin{align}
C_{sym}(\sigma) &=  -\int_{-\infty}^{\infty} P(y) \log_2 P(y) dy \nonumber\\
+&\frac{1}{2} \int_{-\infty}^{\infty} L[y|0] \log_2 L[y|0] dy 
+ \frac{1}{4} \log_2 (2 \pi \sigma^2 e), 
\end{align}
denotes the symmetric information rate of the two-way relay channel under the IID assumption
and $P(y)  \!=\! (1/2) L[y|0] \!+\! (1/2) L[y|1]$ is the PDF of
a received symbol under the assumption.

The BP thresholds of various regular LDPC ensembles versus the code rate are shown in Fig.~\ref{zu_s1}.
We search BP thresholds by evaluating BER using
the population dynamics DE with $N\!=\!10^5$ and $T\!=\!2000$.
It is confirmed that the estimation is accurate up to the third decimal place.
The BP thresholds have a gap to symmetric information rate as predicted in~\cite{Sula17}.

\subsection{Spatial coupling coding for two-way relay channels}

We now turn to SC-LDPC codes.
In this paper, we examine the simplest $(d_l,d_r,L)$-LDPC codes with chain length $L$
where $k=d_r/d_l$ and $\hat{d}_l=(d_l-1)/2$ are integers.
The protograph is then uniquely defined~\cite{Kudekar11}, which makes the structure of the population dynamics DE relatively simple.
The DE analysis for general protograph codes is left as open here.

\begin{figure}[!t]
\centering
\includegraphics[width=0.75\linewidth]{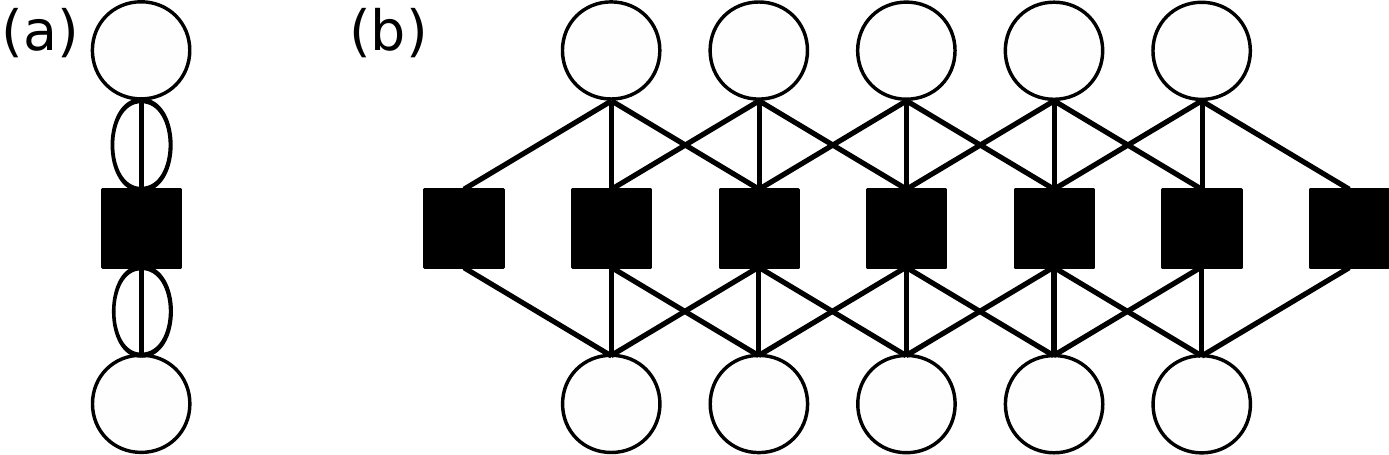}
\caption{A protograph of (a) $(3,6)$-LDPC codes and (b) $(3,6,5)$-LDPC codes.
}\label{zu_s2}
\end{figure}

A protograph of $(d_l,kd_l)$-LDPC codes is represented by $k$ variable nodes and one check node,
e.g., (a) of Fig.~\ref{zu_s2}. 
To construct a protograph of SC-LDPC codes, we prepare $L$ copies of the protograph of an uncoupled code and 
attach $\hat{d}_l$ check nodes to each side of copies.
Edges of the protograph are then assigned from a variable node
to check nodes within ``distance'' $\hat{d}_l$, e.g., (b) of Fig.~\ref{zu_s2} where $(d_l,d_r,L)=(3,6,5)$.
As a result, we have $L$ bundles of $k$ variable nodes labeled by $i\in [L]$, and
 $L+2\hat{d}_l$ check nodes labeled by $a\in\{-\hat{d}_l+1,\dots,L+\hat{d}_l\}$, where
 check nodes labeled from $1$ to $L$ are derived from original protographs.
The design rate is given by $1-(L+2\hat{d}_l)/(kL)$, which recovers that of uncoupled codes as $L\rightarrow\infty$.
 
Let us evaluate BP thresholds for SC-LDPC codes with finite $L$. 
In a protograph, each variable and check nodes respectively have
a PDF $P^{(l)}(m|z)$ and $Q^{(l)}(\hat{m}|z)$ of messages as in the last subsection.
Those PDFs are propagated as messages on a protograph.
From a symmetric structure in each bundle,
$P^{(l)}_{i\rightarrow a}(m|z)$ denotes the PDF of message $m$ as a message
 from a variable node in the $i$-th bundle 
 to a check node $a$ at the $l$-th step.
Similarly, let us denote the PDF of message $\hat{m}$ as a message
 from a check node $a$ to a variable node in the $i$-th bundle 
  by $Q^{(l)}_{a\rightarrow i}(\hat{m}|z)$.
DE equations of binary two-way relay channels and $(d_l,d_r,L)$-LDPC codes then read 
\begin{align}
&P^{(l)}_{i\rightarrow a}(m|z)=
\int dyL[y|z]\int \prod_{b\in N(i)\backslash a}d\hat{m}_bQ^{(l-1)}_{b\rightarrow i}(\hat{m}_b|z)\nonumber\\
&\mathrel{\phantom{P^{(l)}_{i\rightarrow a}(m|z}}\times\delta\left(m-\lambda(y)-\sum_{b\in N(i)\backslash a}\hat{m}_b\right), \label{eq_n3}\\
&Q^{(l)}_{a\rightarrow i}(\hat{m}|z)
=\frac{1}{2^{d_r-2}}\sum_{\{z^{(s)}_j\}\in S'}
\int \prod_{s=1}^{k-1}d{m}^{(s)}_iP_{i\rightarrow a}^{(l)}({m}^{(s)}_i|z^{(s)}_i)\nonumber\\
&\times \prod_{j\in N(a)\backslash i}\left(\prod_{s=1}^kd{m}^{(s)}_jP_{j\rightarrow a}^{(l)}({m}^{(s)}_j|z^{(s)}_j)\right)\nonumber\\
&\times\delta\left(\hat{m}-2\tanh^{-1}\left[
\prod_{(j,s)\neq (i,k)} \tanh\left(\frac{{m}^{(s)}_j}{2}\right)\right]\right), \label{eq_n4}
\end{align}
where $N(\cdot)$ is a set of neighboring nodes in a protograph and
 $S'\!\triangleq\!\left\{\{z^{(s)}_j\}_{j\in N(a)}^{s\in [k]}\!\in\! \{0,1\}^{d_r};\bigoplus_{j,s}z^{(s)}_j\!=\!0, z_{i}^{(k)}\!=\!z\right\}$.
A protograph of uncoupled LDPC codes recovers~(\ref{eq_d3}) and~(\ref{eq_d4}).

\begin{figure}[!t]
\centering
\includegraphics[width=0.87\linewidth]{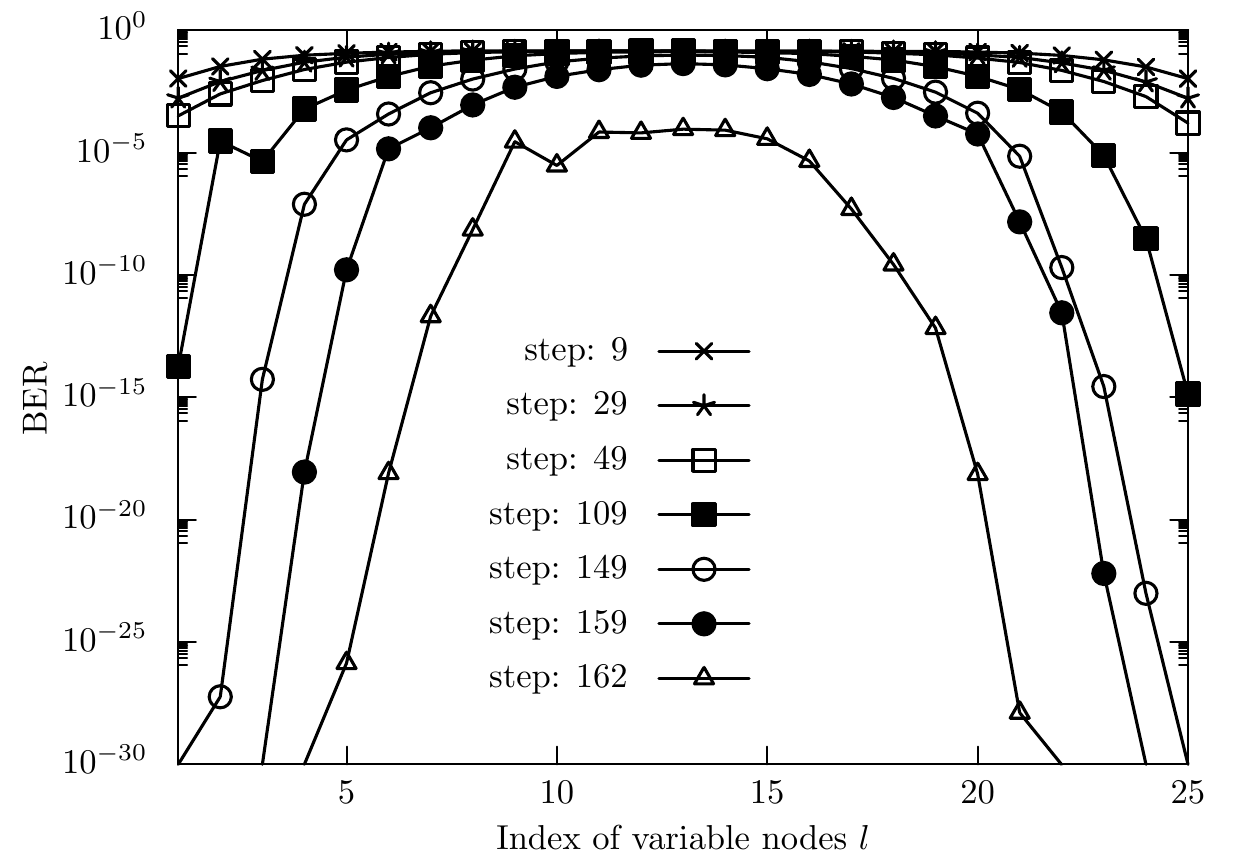}
\caption{BER of each variable node in $(3,6,25)$-LDPC codes with several DE steps evaluated by DE equations for a two-way-relay channel.
}\label{zu_s3}
\end{figure}

Population dynamics is implemented as an extension of Algorithm 1.
In this case, we prepare $4Ld_l$ populations with size $N$ 
to approximate PDFs $P^{(l)}_{i\rightarrow a}(\cdot|z)$ and $Q^{(l)}_{a\rightarrow i}(\cdot|z)$.
Fig.~\ref{zu_s3} shows dynamics of BER of each variable node in $(3,6,25)$-LDPC codes
when $N=10^4$ and $\sigma=0.78$.
It is apparent that they decrease from each side of the chain, as observed in the symmetric channel case~\cite{Kudekar11}.
BERs vanish after the 169th step indicating that the code is decodable.

\begin{figure}[!t]
\centering
\includegraphics[width=0.87\linewidth]{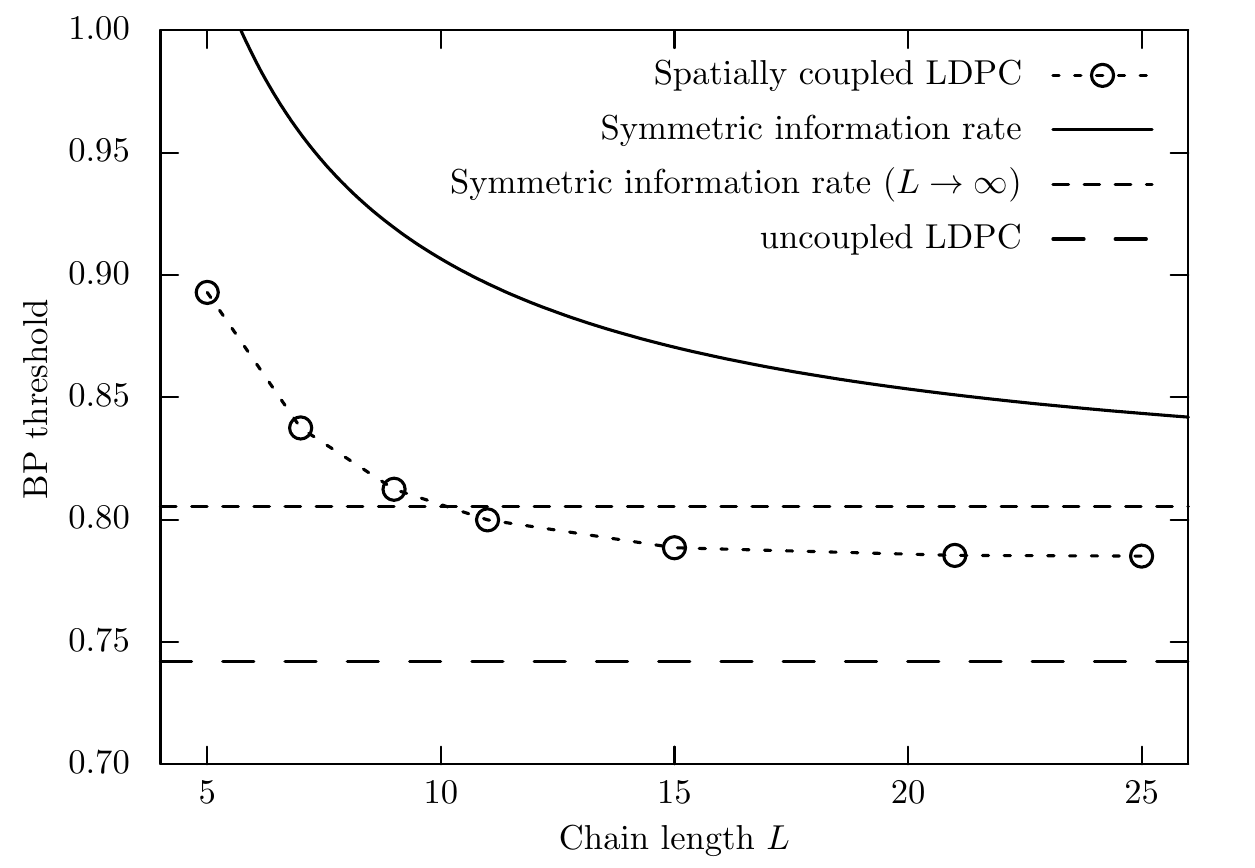}
\caption{BP threshold for a two-way-relay channel and $(3,6,L)$-LDPC codes as a function of chain length $L$.
The long-dashed line represents the BP threshold of uncoupled $(3,6)$-LDPC codes.
The solid line and short-dashed line respectively represent the symmetric information rate of the SC-LDPC codes and its limiting value as $L\rightarrow\infty$.
}\label{zu_s4}
\end{figure}

Fig.~\ref{zu_s4} shows the BP threshold of $(3,6,L)$-LDPC codes and symmetric information rate corresponding to the design rate.
In population dynamics, we use $N\!=\!10^5$ and $T\!=\!2000$.
The results indicate that the BP threshold is monotonously decreasing as $L$ increases.
The limiting value is estimated as $0.785$ by extrapolation,
which lies between the BP threshold $0.742$ of  the uncoupled $(3,6)$-LDPC codes 
and the correspondent symmetric information rate $0.805$.
The same is true for $(3,9,L)$-LDPC codes: the spatially coupling coding achieves $0.647$ ($L\rightarrow \infty$) 
while the BP threshold and the symmetric information rate
of uncoupled codes are respectively given by $0.624$ and $0.666$.
It is noteworthy that our evaluation underestimates BP thresholds because $T\!=\!2000$ is not sufficient in general. 
It is known that a BP decoder for spatially coupled codes needs a large number
of iterations before convergence~\cite{Kudekar11} especially around the threshold.
These facts suggest that the spatial coupling coding successfully improves BP thresholds
although whether it achieves the MAP threshold
 or not is still left to open.

\section{Summary}
In this paper, asymptotic behavior of LDPC codes and SC-LDPC codes for two-way relay channels are studied.
Combining the population dynamics DE with DE formulas for asymmetric channels, BP thresholds of regular LDPC codes 
are evaluated.
In addition, we provide the DE equations of $(d_l,d_r,L)$-LDPC codes and performed the population dynamics DE.
The results show that the spatial coupling coding successfully improves the BP thresholds of two-way relay channels.

\section*{Acknowledgement}

This work is supported
by JSPS Grant-in-Aid for Scientific Research (A) Grant Number 17H01280.

\end{document}